\documentstyle[prd,aps,multicol,amsmath,amsfonts,amssymb,epsfig,psfrag]{revtex}

\begin{document}

\title{Quantum optical weak measurements can visualize photon
dynamics in real time}

\author{J\"urgen Audretsch, Thomas Konrad and Artur Scherer}
\address{Fakult\"at f\"ur Physik der Universit\"at Konstanz\\
Postfach M 673, D-78457 Konstanz, Germany}
\maketitle

\begin{abstract}
An experiment is proposed to visualize stroboscopically in
real time the dynamics of a photon oscillating between  two cavities. The
visualization is implemented by a sequence of weak measurements (POVM),
which are carried out by probing one of the cavities with a 
Rydberg atom and detecting 
a  resulting phase shift by Ramsey
interferometry. This way to measure the number of photons 
 in a cavity
was experimentally realized by Brune et al.\ . We suggest a feedback
mechanism which minimizes the disturbance due to the measurement and 
enables a detection of the original evolution of the radiation field.
\end{abstract}
\begin{multicols}{2}
There is much experimental progress in trapping single atoms, 
ions and photons. It has recently been reported that it is possible 
to trap individual atoms with a single photon in a cavity and to 
reconstruct the trajectory of the atom 
\cite{Hood.et.al00Pinkse.et.al00}.
We want to contribute to this rapidly growing field of quantum 
visualization by proposing an experimental realization in a very 
\lq\lq clean'' setting. 
In a preceeding paper \cite{AudretschKonradScherer00} 
we have shown theoretically and numerically that it is 
feasible to monitor in real time a dynamical process 
occurring in an individual two-level system with state vector 
\begin{equation}
|\tilde{\psi}(t)\rangle=\tilde{c}_1(t)|\varphi_1\rangle+ 
\tilde{c}_2(t)|\varphi_2\rangle\;.
\end{equation}
Our aim in the following is to describe an experimental 
set up for quantum visualization which serves to register the time behavior of $|\tilde{c}_2(t)|^2$ approximately, 
while only weakly influencing the original dynamics of
$|\tilde{\psi}(t)\rangle$. Here we assume the original dynamics to be
known before the measurements. Why should we want to measure a
theoretically known dynamics? Because we want to verify that our
measurement procedure works. It is a non-trivial task to measure
$|\tilde{c}_2(t)|^2$ -- a quantity which refers to an ensemble -- by a
one run measurement on a single system in real-time. In order to
achieve this goal we need  only  little a priori information
about the original dynamics. In the concrete example of Rabi
oscillations presented in this paper we succeed  with the proposed
measurement scheme knowing only the order
of magnitude of the oscillation period. 
Our long-term objective
is to devise a measurement scheme for the real time monitoring of
partly unknown dynamics.  

In classical physics it is possible to track the evolution of an 
individual system without disturbing it substantially.
How can this aim at least approximately be achieved for quantum
systems? Since projection measurements,  which are also called sharp
measurements, severely alter the original  motion of
$|\tilde{\psi}(t)\rangle$, they are not suitable in an one  shot situation
where only  a single realization of this motion is available. 
An exception are QND schemes. They have the disadvantage to require
for non-trivial dynamics  an observable with a continuous spectrum
\cite{Peres89}, which does not exist for two-level systems. In case of
a two-level system one needs instead  weak (or unsharp) measurements
(POVM measurements) by which the state of the
system is less  disturbed but nevertheless some information about the state  
is provided. 
A single weak measurement can be realized by 
suitably entangling the two-level system with a quantum meter via a  
unitary transformation (premeasurement) followed by a projection measurement 
on the meter. The latter supplies a measurement result, which is read off.

To track the development of $|\tilde{c}_2(t)|^2$  in 
time, a sequence of weak measurements is necessary. 
The corresponding series of measurement results  can
then be appropriately processed  in real time
to give the final measurement readout.
Two conditions
may thus be fulfilled  simultaneously:  i.) The back action of the 
measurements does only
moderately  disturb the original dynamics of the system given by the evolution of
$|\tilde{\psi}(t)\rangle$   and  ii.) the variance of the measurement
results is small
enough to enable a reliable estimate of the original time behavior  of
$|\tilde{c}_2(t)|^2$.  Of course there is no information gain without
disturbance of the state. In fact the greater the information gain the
greater is the change of the state due to the measurement. 


In this paper we sketch an experiment to visualize Rabi oscillations 
with frequency $\Omega_R$. When fulfilling the conditions i.) and ii.) 
we have some freedom in adjusting the measuring apparatus. For example 
$\Omega_R$ does not have to be known exactly beforehand. An apparatus 
tuned to any frequency out of the interval
$[0.75\,\Omega_R\,,\,1.5\,\Omega_R] $  would reveal $\Omega_R$ as the actual
frequency (see below).

In cases in which a 
sequential measurement may be approximately treated as  a continuous measurement it can
in the selective regime be described by a stochastic master equation
\cite{Wiseman93}. This powerful calculational tool enables an optimal exploitation of the information
contained in the readout of the continuous
measurement  \cite{Korotkov00Doherty.et.al00}.

We are considering instead a series of well separated weak measurements.
Such a series  was employed to carry out a QND measurement 
of  small photon numbers in an experiment of Brune, Haroche 
et al.\  \cite{Brune.et.al90}, which was theoretically analyzed  in
\cite{BruneHarocheRaimond92SchackBreitenbachSchenzle92} and  
experimentally realized in \cite{Brune.et.al94}. While the 
weakness of the measurements has been considered as an obstacle 
there, it turns out to be an advantage when it comes to the 
detection of dynamics. The experimental setup we sketch 
in the following is based on the Brune-Haroche experiment. 
We have added a second cavity and a feedback mechanism. 
The system consists of one photon with frequency $\omega $ shared by two  
equally constructed, coupled cavities ${\cal C}_1$ and ${\cal C}_2$. 
One could think of two identical cavities connected by a
waveguide (cf.\ \cite{Skarjaetal99RaimondBruneHaroche97}) or  
 a transmissive mirror. For calculations the
cavities are assumed to have infinite damping time. A justification will
be given bellow. Their coupling is modeled by the interaction Hamiltonian of
the  Jaynes-Cummings type:
\begin{equation}
\label{Ham_Sys}
H=\hbar g(a_1 a^{\dagger}_2+a^{\dagger}_1 a_2)\,
\end{equation}
with coupling constant $g$.
In the interaction picture, which we are
going to use, (\ref {Ham_Sys}) is the full Hamiltonian. The indices 
refer to the cavity numbers. Such a coupling of two cavities
has also been considered in \cite{ZoubiOrenstienRon00}.
The photon which is delocalized over the two cavities can be described as 
a superposition of two states:
\begin{equation}
\label{State_Sys}
|\tilde{\psi}(t)\rangle=\tilde{c}_1(t)\,|1,0\rangle+
\tilde{c}_2(t)\,|0,1\rangle\,.
\end{equation}
The first and the second slot in the ket represent the number of
photons in cavity ${\cal C}_1$ and cavity
${\cal C}_2$ respectively. For the initial state
$|\tilde{\psi}(t=0)\rangle = |1,0\rangle $ we find 
Rabi-oscillations with the Rabi-frequency $\Omega_{\mbox{\scriptsize
$R$}}:=2g$ 
\begin{equation} 
|\tilde{c}_2(t)|^2=\sin^2(gt)\;.
\end{equation}

\begin{figure}[htb]
\centering
\epsfig{file=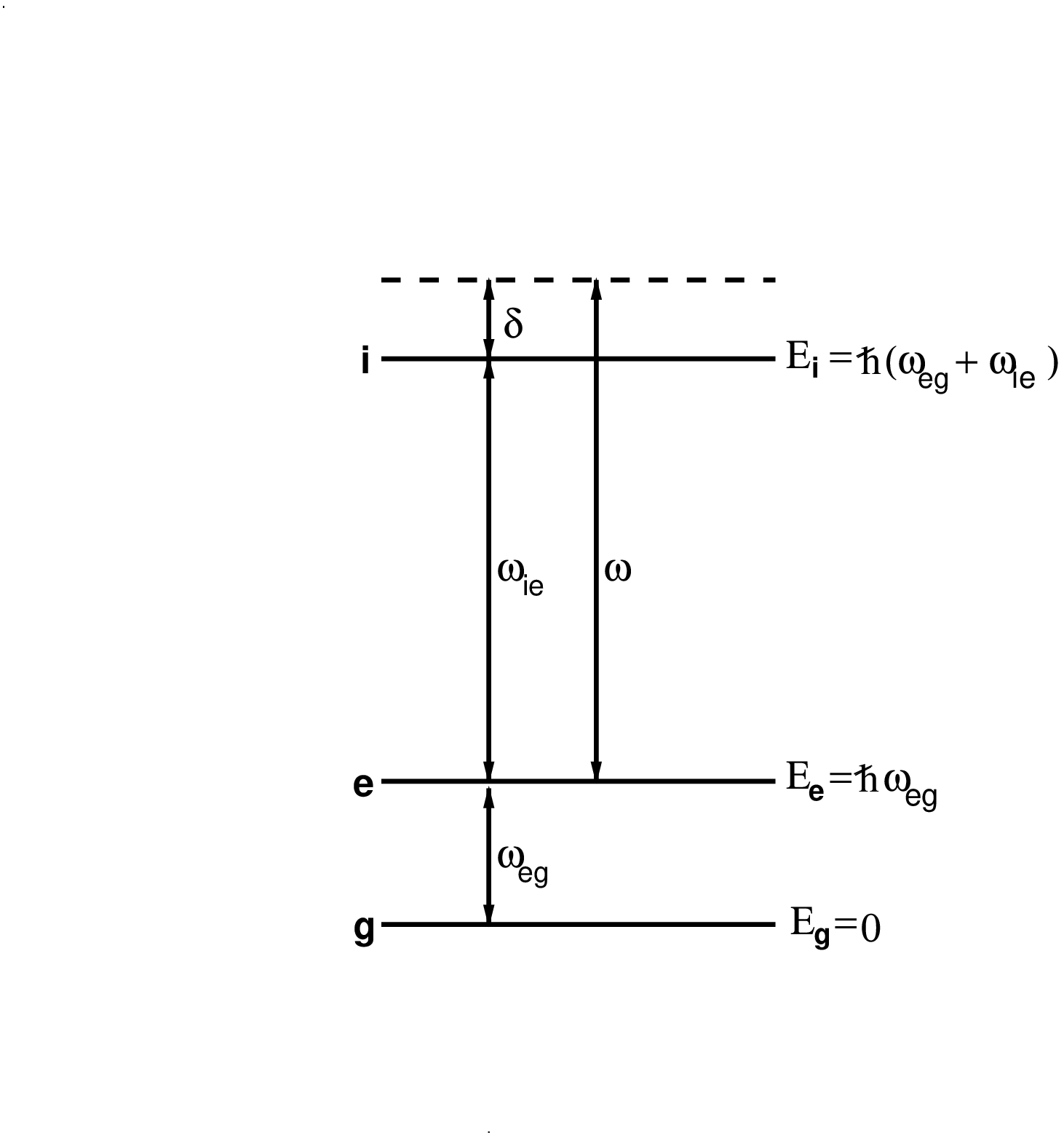,width=0.6\linewidth}
\caption{Relevant levels of Rydberg atom\label{figure1b} }    
\end{figure}

It is our goal  to measure this original  
evolution of $|\tilde{c}_2(t)|^2$ in real time by probing the coupled
cavities with atoms. We sent a  Rydberg atom with three effective energy levels
$g$, $e$, $i$ (see Fig.\ref{figure1b}) and velocity $v$ through the first cavity  ${\cal C}_1$ (cp.
\cite{Brune.et.al90,BruneHarocheRaimond92SchackBreitenbachSchenzle92}). The passage time 
$L_{\mbox{\tiny${\cal C}$}}/v$ ( $L_{\mbox{\tiny${\cal C}$}}$ is the cavity length) is assumed to be much shorter than the period
$T_{\mbox{\scriptsize $R$}}:=2\pi/ \Omega_{\mbox{\scriptsize
$R$}}=\pi/g$ of the oscillations of
$|\tilde{c}_2(t)|^2$. Then the coupling of the two cavities is negligible
during the time the atom spends in the cavity.  The detuning of the atomic
transitions with respect to the frequency of the cavity mode $\omega $ is such
that the interaction
between the atom and ${\cal C}_1$ is dispersive and only the energy
levels $e$ and $i$ suffer an appreciable dynamical stark shift. Provided the
atom enters the cavity in in a superposition of states $|g\rangle$ and
$|e\rangle$, the effective Hamiltonian reads (cp.\ eqn.\ (16) in
\cite{BruneHarocheRaimond92SchackBreitenbachSchenzle92}):  \begin{equation}
H_{\mbox{\scriptsize int}}=\frac{\hbar \Omega^2}{\delta}\,|e\rangle \langle
e|\otimes   a^{\dagger}_1 a_1\,, \label{fieldatomint} \end{equation}
where $\delta:= \omega -\omega_{ie}$ and $\Omega=
\overline{\Omega(r)}$ is the Rabi
frequency averaged over the path  of the atom through the cavity.
With (\ref{fieldatomint}) the state of the enlarged system
composed of the atom and the photon field changes according to
\begin{eqnarray}
\left(c_e|e\rangle+c_g|g\rangle\right)\otimes|\psi\rangle&\rightarrow &
c_e |e\rangle\otimes U_{{\cal C}_1}| \psi \rangle\nonumber\\ 
&&+c_g|g\rangle\otimes | \psi \rangle
\end{eqnarray}
with $U_{{\cal C}_1}$ being diagonal in the basis
$|1,0\rangle$ and $|0,1\rangle$: 
\begin{equation}
\label{relphase}
U_{{\cal C}_1}:=e^{-i\varepsilon_1}|1,0 \rangle \langle 1,0|+|0,1 \rangle \langle 0,1|\;,
\end{equation}
and  $\varepsilon_1:=\frac{\Omega^2}{\delta}\frac{L_{\mbox{\tiny
${\cal C}$}}}{v}$. $|\psi\rangle$ represents the state of the 1-photon-field 
probed by atoms. 
The net effect of the atom-field coupling described by
the interaction Hamiltonian (\ref{fieldatomint}) is
that only  the amplitude of the alternative $|e\rangle\otimes|1,0\rangle$
suffers a phase shift $e^{-i\varepsilon_1}$  while the amplitudes
of the other quantum alternatives remain unchanged. 
\begin{figure}[tb]
\centering
\epsfig{file=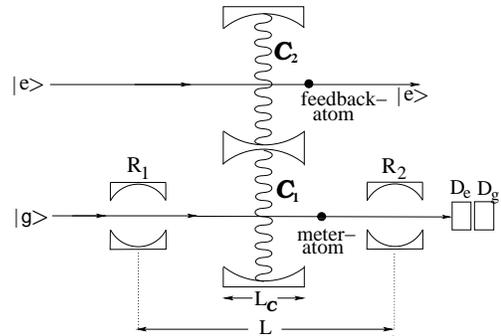,width=0.75\linewidth}
\caption{Experimental setup\label{figure1} }    
\end{figure}

Phase shifts between several quantum
alternatives may be measured by interferometry. As proposed in
\cite{Brune.et.al90} it is convenient to use the Ramsey method of
separated oscillatory 
fields (see Fig.\ \ref{figure1}).   
To this end a Rydberg atom  is initially prepared in state
$|g\rangle$. Before entering cavity ${\cal C}_1$ the state of the atom
is  transformed into a 
superposition of states $|e\rangle$ and $|g\rangle$  by entering a
cavity  which contains a first classical
oscillatory microwave field $R_1$ with frequency $\omega_{\mbox{\scriptsize
$r$}}$. In the cavity ${\cal C}_1$ the atomic state becomes entangled with the
state of the cavities as discussed above.  After leaving ${\cal C}_1$ the atom
crosses a second cavity with a classical microwave field $R_2$
which is in phase with $R_1$ and positioned at the distance $L$ from it. The total state
change of duration $\delta \tau$ amounts to (cf.\
\cite{BruneHarocheRaimond92SchackBreitenbachSchenzle92}) 
$|\Psi(t_0)\rangle\rightarrow |\Psi(t_0+\delta \tau)\rangle$, where 
the product state before the atom enters ${\cal C}_1$ is given by 
\begin{equation}
|\Psi(t_0)\rangle=|g\rangle\otimes|\psi(t_0)\rangle=|g\rangle\otimes\big(c_1(t_0)| 1, 0\rangle +c_2(t_0) |0,1 \rangle \big)
\end{equation}
and the final entangled state equals 
\begin{eqnarray}
|\Psi(t_0+\delta\tau)\rangle&=&
|e\rangle\otimes\big(u_1^{e}c_1(t_0)| 1, 0\rangle +
u_2^{e}c_2(t_0)| 0, 1\rangle \big)
\nonumber\\
&+&|g\rangle\otimes\big(u_1^{g}c_1(t_0)| 1, 0\rangle +u_2^{g}c_2(t_0)| 0, 1\rangle \big)\,.\nonumber\\ &&
\label{entangled}
\end{eqnarray}
The coefficients in
(\ref{entangled}) are given by
\begin{eqnarray}
u_1^{e}&=&\frac{1}{2}\sin\left(\frac{\pi}{2}\frac{v_0}{v}\right)
\left[e^{i(\varphi_0-\varepsilon)\frac{v_0}{v}}+1 \right]\label{amplitudes1}\\
u_2^{e}&=&\frac{1}{2}\sin\left(\frac{\pi}{2}\frac{v_0}{v}\right)
\left[e^{i\varphi_0\frac{v_0}{v}}+1 \right]\nonumber\\
u_1^{g}&=&\cos^2\left(\frac{\pi}{4}\frac{v_0}{v}\right)-
\sin^2\left(\frac{\pi}{4}\frac{v_0}{v}\right)
e^{i(\varphi_0-\varepsilon)\frac{v_0}{v}}\nonumber\\
u_2^{g}&=&\cos^2\left(\frac{\pi}{4}\frac{v_0}{v}\right)-
\sin^2\left(\frac{\pi}{4}\frac{v_0}{v}\right)
e^{i\varphi_0\frac{v_0}{v}}\nonumber
\end{eqnarray}
with $\varepsilon= 
\frac{\Omega^2}{\delta}\frac{L_{\mbox{\tiny ${\cal C}$}}}{v_0}$. $v_0$
characterizes the Ramsey fields and depends on the length $l_{\mbox{\scriptsize $r$ }}$ of
each of the corresponding cavities and the effective Rabi-frequency
$\Omega_{\mbox{\scriptsize $r$ }}$ inside these cavities: $v_0:=2l_{\mbox{\scriptsize $r$
}}\Omega_{\mbox{\scriptsize $r$ }}/\pi$.
$\varphi_0:=(\omega_{\mbox{\scriptsize $r$ }}-\omega_{eg})
\frac{L}{v_0}$ is the phase shift which is induced by the Ramsey cavities in
the case  $v=v_0$.
An analogous result was obtained in eqn.\ (A7) of \cite{BruneHarocheRaimond92SchackBreitenbachSchenzle92}
for the initial atomic state being $|e\rangle$. Eqn.\
(\ref{entangled}) shows that the meter states $|e\rangle$ and
$|g\rangle$ couple in general to both cavity states $| 1, 0\rangle$ and
$| 0, 1\rangle$. This is a characteristic trait of a weak measurement.      

After the atom has left the second Ramsey field $R_2$ its energy is 
finally detected in a projection measurement by field ionization counters $D_e$ and $D_g$. 
The state of the composite system after a measurement with result
$l\in\{e,g\}$  reads
$|\Psi_l(t_0+\delta \tau)\rangle=| l \rangle\otimes|\psi_l(t_0+\delta
\tau)\rangle $ with photon state \begin{eqnarray}
\label{change}
|\psi_l(t_0+\delta\tau)\rangle &=&
|u_1^{l}|\,c_1(t_0)| 1, 0\rangle \\
&&+|u_2^{l}|\,e^{i(\chi^l_2-\chi^l_1)}c_2(t_0)| 0, 1\rangle \,.\nonumber
\end{eqnarray}
and 
$u^l_j=|u^l_j|e^{i\chi^l_j}$ for $j\in\{1,2\}$.
Here a global phase factor has been omitted.
The probability to obtain the related measurement result $l$ is given by  the
expectation value of the corresponding projector: 
$\mbox{prob}(l)=\langle\,\left(\hspace{0.5mm}|l\rangle
\langle l|\otimes  1\hspace{-1.41mm}1
 \right)\,\rangle_{\Psi(t_0+\delta\tau)}$. Eqn.\ (\ref{change}) shows that
after the measurement the photon is in general not localized in one of the
cavities. The disturbance of the photon state due to the measurement may be
small. Because of the Rabi-evolution between the measurements this set up 
represents no QND measurement of the photon number as it has been in
the Brune-Haroche experiment.

Referring to the photon field only, the change of its state due to a
single measurement with result $l$ can be  expressed by an operation
$M_l$:
$|\psi(t_0)\rangle \rightarrow
|\psi_l(t_0+\delta\tau)\rangle=M_l|\psi(t_0)\rangle$. Like all bounded
operators, $M_l$ can be written as \lq\lq phase`` times \lq\lq modulus``
(polar decomposition) \begin{equation}
M_l=U_l|M_l|
\label{operation}
\end{equation}
with unitary transformation 
\begin{equation}
U_l:=| 1, 0\rangle \langle 1,0|+e^{i(\chi^l_2-\chi^l_1)}| 0, 1\rangle
\langle 0,1|\label{U_l}
\end{equation}
and positive operator 
\begin{equation}
|M_l|:=|u_1^{l}|\,| 1, 0\rangle \langle 1,0|+|u_2^{l}|\,| 0, 1\rangle \langle 0,1|\;.
\end{equation}
The probability to obtain the outcome  $l$ is then: 
\begin{equation}
\mbox{prob}(l)=\langle\, M^{\dagger}_lM_l\,\rangle_{\psi(t_0)}=\langle\,
|M_l|^2\,\rangle_{\psi(t_0)}\;. \label{prob}
\end{equation}
The $E_l:=|M_l|^2$ is also called effect. In this way we obtain e.g. for the
probability to measure the energy e: $p_e=p_1|c_1|^2+p_2|c_2|^2$, where
$p_j:=|u_j^{e}|^2$ is fixed by (\ref{amplitudes1}). 

The effects have the property
$E_e+E_g=1\hspace{-1.41mm}1$ and generate a positive
operator valued measure (POVM). In the special case where $u_1^e=u_2^g=1$
and $u_2^e=u_1^g=0$, 
the operation $M_l=E_l$ is a projector. If 
on the other hand $E_l=1\hspace{-1.41mm}1$, no measurement at all has taken place but only an
unitary development ($M_l=U_l$). These two cases are the extremes of a
sharp and a 
totally unsharp measurement. By varying the parameters 
$v\,,v_0\,,\varphi_0$ and $\varepsilon$ of the setup all degrees of \lq\lq
weakness'' between 
these two extreme cases as well as the extremes  themselves can be reached. 

Eqn.\ (\ref{prob}) shows that the information obtained by the
generalized measurement is solely contained in  $|M_l|$. This part of
the operation $M_l$  in (\ref{operation}) represents at the same time the
unavoidable minimal disturbance of the system by the measurement. 
But our set up  causes in addition by means of $U_l$ a purely unitary or
Hamiltonian evolution of the state, which modifies the photon  state without
being necessary for the extraction of information. Since we want to disturb
the original state motion as little as possible, we have to
install a Hamiltonian feedback mechanism which compensates $U_l$ given
by (\ref{U_l}). Such a procedure has already been proposed by Wiseman 
\cite{Wiseman95}. In our case feedback can be implemented by modifying 
the set up as follows:

After a 
measurement beginning at an arbitrary time $t=t_0$ with outcome $l$ an atom prepared in state
$|e\rangle$ is sent through the second
cavity ${\cal C}_2$. As in the case where an atom crosses  cavity
${\cal C}_1$ the unitary evolution is again governed by the dynamical Stark
effect, with the only difference that now the energy shift depends on the
number of photons in ${\cal C}_2$ instead of ${\cal C}_1$:
\begin{equation}
|e\rangle\otimes|\psi_l(t_0+\delta \tau )\rangle\rightarrow 
|e\rangle\otimes U_{{\cal C}_2}|\psi_l(t_0+\delta \tau )\rangle\,, 
\end{equation}
with 
\begin{equation}
U_{{\cal C}_2}:=|1,0 \rangle \langle 1,0|+e^{-i\varepsilon_2}|0,1 \rangle
\langle 0,1|\;. \end{equation}   
The combined influence of the measurement and feedback leads to
\begin{equation}
U_{{\cal C}_2}|\psi_l(t_0+\delta \tau )\rangle=U_{{\cal C}_2}U_l|M_l|\,|\psi(t_0)\rangle\;. \end{equation}
\noindent The condition for compensation of $U_l$ in (\ref{U_l}) is
therefore $U_{{\cal
C}_2}U_l=1\hspace{-1.41mm}1\,\Leftrightarrow\,\varepsilon_2=\chi^l_2-\chi^l_1$,
where $\chi_j^l$ may be obtained from (\ref{amplitudes1}). This condition demands that the compensating phase
$\varepsilon_2=\frac{\Omega^2}{\delta_f}\frac{L_{\mbox{\tiny  ${\cal
C}$}}}{v_f}$ ($f$ denotes the feedback) has to be chosen depending  on
the measurement outcome $l$. We see two ways to vary 
$\varepsilon_2$. One is to select an appropriate velocity $v_f$ of the
feedback atom sent through the upper cavity ${\cal C}_2$.  The other
possibility consists in setting up a suitable detuning $\delta_f$.
This can be done by shifting the atomic energies by means of an static
electric field in the cavity ${\cal C}_2$ cp.\  \cite{TombesiVitaliRaimond00}. 
Please note that it makes no difference whether the atom sent through
cavity ${\cal C}_2$ is thereafter measured or not because the composite system after the
interaction is in a product state. 

In order to reach our final aim of monitoring the original Rabi-oscillations of
$|\tilde{c}_2(t)|^2$, a sequence of measurements at times
$t_n=n\tau$ with $n= 1,2,3\ldots$ has
to be carried out. Between two consecutive measurements the system
evolves undisturbed according to the Hamiltonian (\ref{Ham_Sys}). 
The resulting total evolution of the system is given by $c_2(t)$  instead 
of $\tilde{c}_2(t)$. To process the data obtained in the single measurements   
we first of all  divide  the sequence of results with values $e$ and $g$
into groups of $N$. From each so called \lq\lq
$N$-series'' we extract the relative frequency $r:= N_e/N$ of the 
number $N_e$ of $e$-results.  It turns out
\cite{AudretschKonradScherer00} that its  expectation value
${\cal E}(r)$ is related to the value which $|c_2|^2$ assumes immediately
before the start of the $N$-series by 
\begin{equation}
\label{c2n}
|c_2|^2 = \frac{{\cal{E}}(r)-p_1}{\Delta p}
\end{equation}
with $\Delta p:=p_2-p_1=|u_2^{e}|^2 -|u_1^{e}|^2$ of
(\ref{amplitudes1}). 
In a sequence of measurements on a single radiation field we do not
have access to the expectation value ${\cal E}(r)$, which refers to an
ensemble. Instead we insert  $r(t_0)$ into the right hand side of equation
(\ref{c2n}) and obtain thus a \lq\lq best guess''
of $|c_2(t_0)|^2$ at time $t_0$ when the first measurement
of the respective N-series began:
\begin{equation}
\mbox{G}_2(t_0)=\frac{r(t_0)-p_1}{\Delta p}\,.
\end{equation}
The possible values of $\mbox{G}_2(t_0)$
are distributed around $|c_2(t_0)|^2$ and may be
negative. This estimation
of $|c_2(t_0)|^2$ can be  good only  if the duration of the N-series $N\tau$ is
much smaller than the period $T_{\mbox{\scriptsize $R$ }}$ of the oscillations
of the system. The sequence of $\mbox{G}_2$ at various times 
serves as  the final readout of the sequential measurement.  

We have two competing influences on the system: The strength of the 
original dynamics is proportional to $g$ or $\Omega_{\mbox{\scriptsize $R$
}}=2\pi/T_{\mbox{\scriptsize $R$
}}$.  The measurements on the other hand hinder this dynamics the more 
the stronger they are and the quicker they are repeated with frequency
$1/\tau$. The Zeno effect
demonstrates this! A measure for the disturbance due to the
sequence of measurements is the decoherence time $T_D$, the time after which
the off-diagonal elements of the density matrix (in the
$|\varphi_1\rangle$, $|\varphi_2\rangle$ basis)  have decayed to $1/e$ of
their original value. In a subsequent paper we will show that $T_D=
8\frac{p_0(1-p_0)}{(\Delta p)^2}\tau$ with
$p_0:=(p_1+p_2)/2$. In order to have a 
high resolution in the sequential measurement it would be desirable
to have a small decoherence time ($T_D$ is proportional to the time it
takes to distinguish between the states $|\varphi_1\rangle$ and 
$|\varphi_2\rangle$, cp.\cite{AudretschKonradScherer00}). 
On the other hand the influence of the measurement
should not dominate the evolution. We therefore require the
decoherence time to be maximally as great as the Rabi time. A
numerical analysis \cite{AudretschKonradScherer00} showed
 that for our purpose a favorable balance 
of information gain and disturbance is obtained if the so called fuzziness $f:=
\frac{\pi T_D}{2T_{\mbox{\tiny $R$ }}}$  is 
adjusted to be close to one: $f\approx 1$. In fact it suffices to choose 
$f$  in  the interval $0.75\lesssim f\lesssim 1.5$.
The experimental parameters $\varepsilon$, $\varphi_0$, $v_0$,
$v$ and $\tau$  have to be fixed correspondingly.

\begin{figure}[htb]
\centering
\epsfig{file=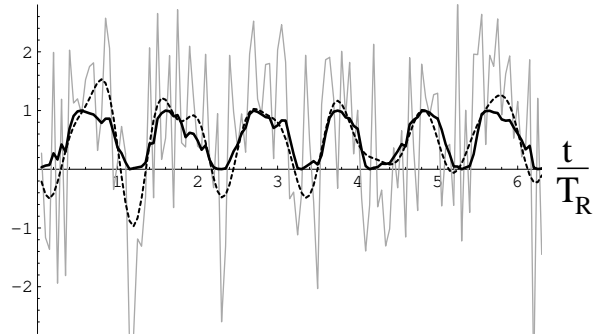,width=0.95\linewidth}
\caption{ \label{figure2}
Under a sequence of appropriate weak measurements the measurement
readout $\mbox{G}_2$ (grey curve) is correlated with the state
evolution $|c_2(t)|^2$ (black curve). This becomes evident after noise
reduction (dashed curve) of the readout $\mbox{G}_2$, which was
carried out taking into account approximately $12$ Rabi-cycles. 
Parameter values: $\frac{v}{v_0}\in[1.25-0.125,1.25+0.125]$,
$\varepsilon=0.068\cdot\pi$, $\varphi_0=\pi$, $\tau=0.002
T_{\mbox{\scriptsize $R$}}$ and $N=25$, which lead to an average 
fuzziness of $f=0.98$.}
\end{figure}
We have simulated numerically all the processes  described above including 
the feedback. In a realistic experiment the velocity $v$ of the probing 
atoms  and the feedback atoms will vary from one single measurement to the other. We have 
accordingly tolerated the velocities $v$ to fluctuate uniformly by 
$\pm 10\%$ about the desired mean value.
The resulting dynamics of the state under the influence of stroboscopically
applied weak measurements is given by the $|c_2(t)|^2$-curve (black) in 
Fig.\ \ref{figure2}. The measurement readout which is defined as best guess 
$\mbox{G}_2(t)$ of  $|c_2(t)|^2$ (grey curve) has been further processed 
to the noise reduced $\mbox{G}_2$-curve (dashed).  The noise reduction
procedure consists  essentially in a time-averaging of the
readout. Details are described in Appendix C of
\cite{AudretschKonradScherer00}.

We find a high correlation 
including the phase between the noise reduced $\mbox{G}_2$-curve and 
the $|c_2|^2$-curve. The actual evolution of the state is therefore well 
monitored in time. The  $|c_2|^2$-curve reflects 
the fact that the original Rabi-oscillations have been disturbed by 
the measurement, though they are only slightly modified. 

\begin{figure}[h!]
\centering
\epsfig{figure=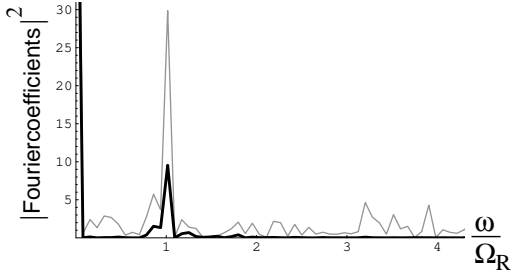,width=0.8\linewidth}
\caption{\label{figure3} Power spectra of $|c_2(t)|^2$ (black) and $\mbox{G}_2(t)$ (grey). }   
\end{figure}

Fig.\ \ref{figure3} 
shows the power spectrum of the  $|c_2|^2$-curve (black) and the measurement 
readout $\mbox{G}_2$ (grey). Both curves  are peaked at the Rabi frequency 
$\Omega_{\mbox{\scriptsize $R$}}$.

All parameters used in Fig.\ \ref{figure2} and \ref{figure3} are
realistic (cf.\ \cite{BruneHarocheRaimond92SchackBreitenbachSchenzle92})  apart
from the number of Rabi cycles   and the
efficiencies of the detector and the feedback mechanism. In what
follows, we are going
to address these remaining problems. 

The number of Rabi cycles which can be monitored depends crucially on
the mean  lifetime $t_{\mbox{cav}}$ of the coupled cavities which has so
far been assumed to be infinite. At the required frequency of $50$ GHz  
 a $t_{\mbox{cav}}= 0.1$ s seems to be feasible  (cf.\
\cite{BruneHarocheRaimond92SchackBreitenbachSchenzle92}). The time of flight of an atom with
velocity $v = 1000$ m/s through cavity ${\cal C}_1$ amounts to $L_C/v=10
\mu$s. We therefore assume that a time $\tau=100 \mu$s between two
consecutive measurements suffices to send a feedback atom through
cavity ${\cal C}_2$ between the measurements. Then $1000$ measurements
can be made within the mean lifetime $t_{\mbox{cav}}$ of the cavities. 
For $T_R/\tau = 500 $ as assumed in Fig.\ \ref{figure2}, this
corresponds to monitoring the first two Rabi cycles.  As can be seen
in Fig.\ \ref{figure2} the readout of
the measurement $\mbox{G}_2$ (grey curve) oscillates about the evolution of
the component $|c_2|^2$ (black curve), so that also for two Rabi
cycles the evolution of $|c_2|^2$  can be approximately recovered by
averaging $\mbox{G}_2$ over an appropriate timescale.

A more serious problem arises when some of the feedback atoms are not
sent through cavity ${\cal C}_2$ and thus the unitary part of the
back-action $U_l$ is not always  
compensated.  In our simulations we obtained still good results
with $4$ percent of the feedback atoms missing while for more than $7$
percent the evolution of  $|c_2|^2$ is significantly disturbed. Since
the preparation of the Rydberg atoms in the Haroche experiment is a
random process, feedback atoms might be missing. This problem could be
solved by sending a high number of atoms at once through  cavity
${\cal C}_2$, each with large detuning, i.e., causing a small
phase shift of the radiation field. The mean number $\bar{n}$ of atoms sent in
one burst should lead to a total phase shift, which compensates $U_l$.
Since the standard deviation divided by the mean number $\Delta n/\bar{n}$
can be made arbitrarily small by increasing $\bar{n}$, $U_l$ can be
compensated very precisely. 

\begin{figure}[t]
\psfrag{t}[B]{$\frac{t}{T_R}$}
\centering
\psfig{file=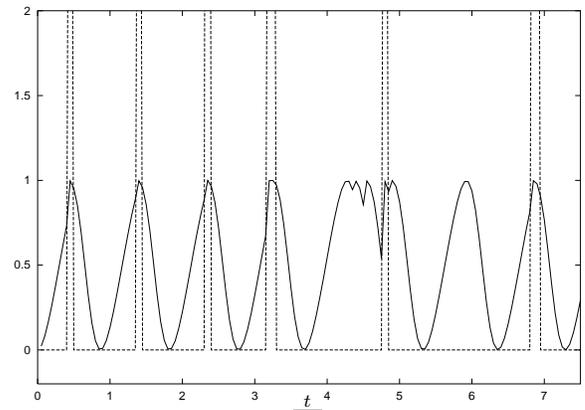,width=0.65\linewidth, angle= -90}
\caption{ \label{figure4}
Simulation of a sequence of weak measurements with Poissonian statistics
and a detector efficiency of $60 \%$.
The back-action of these measurements has no unitary part.  
The dashed  and the black curve represent $\mbox{G}_2$ and $|c_2(t)|^2$
respectively.  
Parameter values: $\frac{v}{v_0}\in[20-2,20+2]$,
$\varepsilon=\pi\frac{\bar{v}}{v_0}$, $\varphi_0=0$, $\tau=
0.002 T_{\mbox{\scriptsize $R$}}$ and $N=25$, which lead to an average 
fuzziness of $f=2.04$.}
\end{figure}

Another crucial point is the detector efficiency. The measurement
results are not so sensitive to the loss of information. It is the
disturbance of the Rabi oscillations as a consequence of not knowing
how to  prepare the feedback atoms which  mainly hinders the visualization of
the photon dynamics. The problem can be
solved by employing a detector with more than $96 \%$ efficiency (see
above). 
If only detectors with moderate efficiencies are available  there is
another solution. It consists in choosing the experimental parameters 
in  (\ref{amplitudes1}) such that the measurement statistics become 
Poissonian:
\begin{equation}
u_1^e= 0, \,u_2^e=\tilde\epsilon,\, u_1^g= 1,\,\,\mbox{and}\,\,u_2^g=\sqrt{1-\tilde\epsilon^2},\, 
\end{equation}
where $\tilde\epsilon = \sin(\pi v_0/2v)$ should be small. 
In this case we do not
need feedback because $U_l=1\hspace{-1.41mm}1$ for both $l=e$ and $l=g$  . 
Most measurement results will be \lq\lq g ", in which case the change
of the state will be of higher order in the small parameter
$\tilde\epsilon$. The result \lq\lq e " is likely only when 
$|c_2|^2\approx 1$ and then the state change is also very small. The
disadvantage of the Poissonian method is  that the 
quality of the monitoring suffers from the Poissonian
statistics. The measurement results essentially indicate only the
parts of the state evolution where $|c_2|^2\approx 1$.
For detector efficiencies higher than $96 \%$ the method with feedback
atoms shows clearly better results. For efficiencies below $96 \%$ it
is preferable to avoid the necessity of feedback
atoms. Fig.\ \ref{figure4} displays a simulation of a sequence of
measurements with Poissonian statistics for a
detector efficiency of $60\%$. Roughly the same percentage 
of the maxima of the Rabi oscillations are indicated by the readout
$\mbox{G}_2$. As in the case with feedback we allowed the velocities
of the Rydberg atoms to fluctuate uniformly by $\pm 10\%$ around the mean value $\bar v$.

To sum up: Using the feedback mechanism for detector efficiencies
higher than $96 \%$ the original Rabi-oscillations are well visualized
in phase and frequency. For moderate  detector efficiencies an
alternative method leads to a moderate visibility of the Rabi-oscillations.

We wish to thank the referee for his suggestion to use groups of atoms
for the feedback mechanism in order to improve its accuracy.

This work has been supported by the Optik Zentrum Konstanz.

\end{multicols}
\end{document}